\documentstyle[12pt]{article}
\begin{document}

\begin{center}
{\large\rm Effective action   }
\end{center}
\begin{center}
{\large\rm  for high energy scattering  in QCD\footnote{ Talk at
the International Symposium on Particle Theory in Wendisch-Rietz and at
the DESY workshop on QCD, September 1993. Based on joint work with L.N.
Lipatov and L.
Szymanowski. Collaboration supported in part by Volkswagen - Stiftung.
}}
\end{center} \vspace{1cm} \begin{center}
{\large\it   R. Kirschner}\\{\large\it Leipzig, Germany }
\end{center}
\vspace{5mm}

We study the behavior of scattering amplitudes in the perturbative Regge
region,
\begin{equation}
s \gg \kappa_{\perp }^2 \gg \mu_{had}^2 ,
\end{equation}
with a large energy squared $s$ and transverse momenta of exchanged
particles $\kappa_{\perp }$ much smaller but large enough compared to
the hadron mass scale $\mu_{had}$ to allow for perturbative expansion.
This asymptotics determines the behaviour of structure functions at
small $x$ or of two-jet events with a large rapidity gap.

In this region the dominant contributions come from multi-particle
states with the momenta $k_i$ of produced particles obeying the
conditions of multi-Regge kinematics. $p_A$ and $p_B$ are the momenta of
incoming particles.
\begin{eqnarray}
  s &=& (\mbox{p}_A + \mbox{p}_B)^2 = 2(\mbox{p}_A \mbox{p}_B) \nonumber \\
  s_i &=& (\mbox{k}_i + \mbox{k}_{i-1})^2 = 2(\mbox{k}_i \mbox{k}_{i-1})
          \qquad i = 1, \ldots , n+1 \nonumber \\
  \mbox{k}_0 &\equiv& \mbox{p}_{A'}, \qquad \mbox{k}_{n+1} \equiv
          \mbox{p}_{B'}, \qquad \mbox{k}_i = \mbox{q}_{i+1} - \mbox{q}_i
          \nonumber  \\
  s &\gg& s_1 \sim s_2 \sim \ldots \sim s_{n+1} \gg |\mbox{q}_1^2| \sim
       |\mbox{q}_2^2| \sim \ldots \sim |\mbox{q}_{n+1}^2| \nonumber \\
  s_1 s_2 &\ldots& s_{n+1} = s \prod^n_{i=1} (-k^2_{\perp i}) \nonumber
  \\
  &\mbox{k}_1\mbox{p}_A& \ll \mbox{k}_2\mbox{p}_A \ll \ldots \ll
  \mbox{k}_n\mbox{p}_A     \nonumber \\
  &\mbox{k}_1\mbox{p}_B& \gg \mbox{k}_2\mbox{p}_B \gg \ldots \gg
  \mbox{k}_n\mbox{p}_B
\end{eqnarray}
The result of the leading logarithmic asymptotics for this asymptotics
is known as the perturbative pomeron \cite{FKL}. In this approximation
we have the exchange of two reggeized gluons interacting via s-channel
intermediate gluons with momenta obeying (2).  With increasing energy
the contributions with the exchange of more than two reggeized gluons
become important. Because these contributions are difficult to
calculate, it has been proposed to use an effective action for treating
them \cite{Leff}.

The effective action describes the leading contribution to high energy
peripheral scattering with intermediate states in all sub-energy
channels
obeying (2). It can be obtained by analyzing the leading contribution of
tree graphs. Here we show that it can be derived directly and more
formally from
the ori\-gi\-nal QCD action by approximate integration over modes of the
gauge and the fermion fields that do not appear as scattering or
exchanged gluons or quarks in the considered peripheral high-energy
scattering. Here we restrict ourselves to the pure gluonic case. The
fermionic terms can be obtained analogously or by supersymmetry.

We decompose all vectors with respect to the (almost ) light-like
momenta of incoming particles $ p_A^{\mu } = { \sqrt {s} \over 2}
(1,0,0,1), \ \ \ p_B^{\mu }= { \sqrt {s} \over 2 } (1,0,0,-1) $,

\begin{equation}
k^{\mu } = {2 \over \sqrt s } (k_+ p_B^{\mu } + k_- p_A^{\mu }) +
\kappa_{\perp }^{\mu }
\end{equation}

and introduce the complex notation for the transverse components, $
\kappa = \kappa_{\perp 1} + i \kappa_{\perp 2}, \kappa^{*} =
\kappa_{\perp 1} - i \kappa_{\perp 2} $.
The derivatives in position space $(x^{\mu } )$ are introduces in such a
way that we have $ \partial_+ x_- = \partial_- x_+ = \partial x =
\partial^{*} x^{*} = 1 $. We keep in mind the conditions of multi-Regge
kinematics, explained above. This kinematics dominates
the s- channel intermediate states in high energy peripheral scattering.

We start with the gauge field action in the gauge $ A_-^{a} =  0 $. $
A_+^{a} $ enters quadratically and can be integrated out exactly
resulting in non-local termes (involving the inverse of the derivative
operator $\partial_- $ ).

\begin{eqnarray}
  S_1 &=& \int d^4x \left\{ - \frac{1}{2} A^a \Box A^{a*} +
       i\frac{g}{2} \left[ A f^a
\stackrel{\leftrightarrow}{\partial}_- A^* \right]
       \cdot \partial^{-1}_- (\partial A^a
      + \partial^* A^{a*})  \right. \nonumber \\
 && \left. - \frac{g^2}{8} \left[ (A^* f^a
 \stackrel{\leftrightarrow}{\partial}_- A) \right] \partial^{-2}_-
 \left[ A f^a
    \stackrel{\leftrightarrow}{\partial}_- A^*  \right] \right.
      \nonumber \\
&&  - i\frac{g}{2} \left[Af^aA^*\right]\left(\partial A -
\partial^* A^* \left) -  \frac{g^2}{8} \left[ Af^a A^* \right]
\left[ Af^aA^* \right] \right\}   \right.
\end{eqnarray}

$A$ is the transverse gauge field in the complex notation, $(A^{*} f^{a}
A) = -i f^{abc} A^{* b} A^{c} $.

Now we replace $A$ by the sum ${\cal A} + A + A_1 $, where ${\cal A}, A,
A_1 $ represent different modes of the original field with the momenta
obeying
\begin{eqnarray}
{\cal A}:       k_+ k_- \ll k k^* \sim \mu^2_\bot \nonumber \\
A_1 :  k_+ k_- \gg k k^* \sim \mu^2_\bot         \nonumber \\
A:   k_+ k_- \sim k k^* \sim \mu^2_\bot \; .
\end{eqnarray}
In the dominant contributions to the considered processes the
modes ${\cal A}$ (dashed line ) describe the exchanged gluons and $A$
(full line ) the scattering and produces gluons. The modes represented
by $A_1$ (double line) will be integrated  out. In the dominant
contribution the vertices in (2) must enter in such a way that the
fields on which ${\partial_- }$ acts describe particles carrying large
momenta $k_-$ and the fields on which $\partial_-^{-1} $ acts describe
particles carrying small $k_-$. Large longitudinal momentum factors in
the denominator have to be cancelled by corresponding factors in the
numerator. The ordering in $k_-$, typical for the multi-Regge
kinematics, is induced by the $\partial_- $ structure of the interaction
terms. This becomes more explicite by introducing the notation
\begin{eqnarray}
{\cal A_+} = -  \partial_-^{-1} (\partial {\cal A} + \partial^{*} {\cal
A^{*}}),
{\cal A_+}^{\prime } = - \partial_-^{-1} (\partial {\cal A}^{\prime } +
\partial^{*} {\cal A^{\prime * }}) \nonumber \\
{\cal A_-} = \partial_-^{-1} (\partial {\cal A} - \partial^{*} {\cal
A^{*}}),
{\cal A_-}^{\prime } = - \partial_-^{-1} (\partial {\cal A}^{\prime } -
\partial^{*} {\cal A^{*}}^{\prime })
\end{eqnarray}
The current of scattering particles which are close in momenta to the
incoming particle $p_A$, $j_- = {i \over 2} (A^{*} f^{a}
\stackrel{\leftrightarrow}{\partial}_- A)$,
couples to ${\cal A_+}$ only. Particles with large $k_-$ are scattered
off the field ${\cal A_+}$ describing the effect of particles with lower
$k_-$. ${\cal A_-}$ plays the analogous role  for the particles with
large $k_+$. ${\cal A}_{\pm }^{\prime }$ practically decouple. It is
consistent to consider ${\cal A_+}$ and ${\cal A_-}$ as independent.
With the mode decomposition the kinetic term becomes
\begin{equation}
    -2 A_1^a \partial_+ \partial_- A_1^{a*} - \frac{1}{2} A^a
    \Box A^{a*} - 2 {\cal A_+}^a \partial \partial^* {\cal A_-}^{a*}
 - 2 {\cal A_+}^{\prime a} \partial \partial^* {\cal A_-}^{\prime a*}
\end{equation}

The triple vertex in (2) involving $\partial_- $ includes the
contribution of scattering of perticles with large $k_-$ in the field
${\cal A_+}$,
\begin{equation}
 {ig \over 2} \left[ (A+A_1)  f^a \stackrel{\leftrightarrow}{\partial}_-
 (A + A_1) \right] {\cal A}^a_+
\end{equation}
and a contribution of particle production from an exchanged gluon,

\begin{equation}
 ig ((A_1 + A) f^a \partial {\cal A}_-
     ) {\cal A}^a_+
\end{equation}
There are also two contributions related to scattering of particles with
large $k_+$,
\begin{eqnarray}
 && \left. \left[ \partial {\cal A}_- f^a (A+A_1)
 \right] \partial_-^{-1} (\partial A^a +
\partial^* A^{a*})  \right.
\end{eqnarray}
and
\begin{equation}
  {\cal A}^a_-
\partial\partial^* \partial^{-2}_-\left[Af^a
\stackrel{\leftrightarrow}{\partial}_- A^* \right]  .
\end{equation}

The main contribution of the quartic term in (2) can be represented as
the result of connecting the vertex (6) with a vertex of the type (9)
by a ${\cal A_+ A_- }$ propagator. Thus the quartic vertex can be by an
additional vertex of type (9). The signs are opposite leading to
cancellation of the type (9) vertices.

We consider $A_1$ as a small fluctuation ${\cal O}(g)$. The leading
contribution of the integral over these modes is obtained by eliminating
in (2)   $A_1$ by its linearized equation of motion. Closed loops of
$A_1$ are not included in our approximation, which takes into account
$s$-channel intermediate states obeying multi-Regge kinematics.

Besides of the vertex (7) particles can be produced  by bremsstrahlung
with an intermediate state of $A_1$ modes (Fig.1 ) . The quartic
contribution with the intermediate $A_1$ can be represented by triple
vertices conected by a ${\cal A_+ A_- }$ propagator. Thus a new triple
vertex can replace this quartic term. Adding this new triple vertex to
the production vertex (7) results in the effective vertex of gluon
production \cite{FKL}.
\begin{equation}
ig J^{a *} \partial^{* -1} A + {\rm c.c.}, \ \ \ \
J^{a *} = \partial^{*} {\cal A_+} f^{a} \partial {\cal A_-}.
\end{equation}

Whereas the vertex (6) describing scattering of particles with large
$k_-$
comes out directly from the original triple vertex in (2) involving
$\partial_- $
\begin{equation}
g j_- {\cal A_+} , \ \ \ \  j_- = {i \over 2} (A^{*}
\stackrel{\leftrightarrow}{\partial}_- A )
\end{equation}
the  vertex describing the scattering of particles with large $k_+$
results after the cancellation of the type (9) terms from a sum of the
contribution (8) and a new triple vertex arrising from resolving a
quartic term with intermediate $A_1 $ modes by a ${\cal A}_+ {\cal A}_-
$ propagator.

\begin{eqnarray}
g j_{L +} {\cal A_-} , \ \ \ \ \ j_{L +} = {i \over 2} (A_L
\stackrel{\leftrightarrow}{\partial}_+  A_L)  \nonumber \\
\partial A_L = - \partial^{*} A^{*} .
\end{eqnarray}

The relation between $A_L $ and $A$ is just the gauge transformation of
the transverse gauge potential from the gauge $ A_- = 0$ to the gauge
$A_+ = 0 $.

It is convenient to introduce the complex scalar field $\phi^{a}$for
describing the scattering gluons.
\begin{equation}
A^{a} = i \partial^{*} \phi^{a} ,\ \ \ \  \ \  A_L^{a *} = i
\partial^{*} \phi^{a}.
\end{equation}
In this way the gauge relation (12) is resolved and the
non-localities
($\partial^{-1} $) in the effective vertices (10) and (12) disappear.

We have shown that the dominant contribution to peripheral high-energy
scattering with
multi-Regge kinematics in the s-channel intermediate states
to the gluonic action is described by replacing the original interaction
vertices in (2) by effective vertices of scattering (11), (12) and
production (10). With the scalar field $\phi^{a} $ the resulting
effective action reads,
\begin{eqnarray}
  S^{eff}_g &=& S_{ksg} + S_{kpg} + S_{sg} + S_{pg} \nonumber \\
  S_{ksg} &=& - \frac{1}{2} \int d^4x \partial \phi^a \Box
      \partial^* \phi^{a*} ,
        S_{kpg} = -2 \int d^4x {\cal A}^a_+ \partial \partial^*
       {\cal A}^a_- \nonumber \\
  S_{sg} &=& - \frac{g}{2} \int d^4x \left\{ j^a_{-} {\cal A}^a_+
       + j^a_{+}  {\cal A}^a_- \right\} \nonumber \\
  S_{pg} &=& g \int d^4x \left\{ J^{a*} \phi^a + J^a \phi^{a*} \right\}
      \nonumber \\
  j_{-} &=& i [ \partial^* \phi f^a \partial_- \partial \phi^* +
      \partial \phi^* f^a \partial_- \partial^* \phi]  \nonumber
      \\
  j_{+} &=& i[ \partial^* \phi^* f^a \partial_+ \partial \phi +
      \partial \phi f^a \partial_+ \partial^* \phi^*]    \nonumber \\
   J^{a*} &=& \partial^* {\cal A}_+ f^a \partial {\cal A}_-
\end{eqnarray}
This result was first obtained in \cite{Leff}.

E. and H. Verlinde \cite{Verlinde} derived an effective action for high
energy scattering in gauge theories. The scaling argument on which their
derivation is based eliminates the transverse components of the field
strength tensor from the action. In our derivation just these terms play
an important role to obtain the effective vertex for gluon production.



\input FEYNMAN


\begin{picture}(51000,30000)

\drawline\fermion[\S\REG](3000,21000)[4500]
\drawline\fermion[\N\REG](2800,21000)[3000]
\drawline\fermion[\N\REG](3200,21000)[3000]
\drawline\fermion[\N\REG](3000,24000)[3000]
\drawline\fermion[\NE\REG](3000,24000)[4200]
\drawline\scalar[\E\REG](3000,21000)[2]
\put(3000,21000){\circle*{400}}
\put(3000,24000){\circle*{400}}
\put(3500,16500){p}
\put(6500,27000){k}
\put(5000,19500){q-k}

\put(9000,21000){+}

\drawline\fermion[\N\REG](12000,21000)[6000]
\drawline\fermion[\S\REG](11800,21000)[3000]
\drawline\fermion[\S\REG](12200,21000)[3000]
\drawline\fermion[\S\REG](12000,18000)[1500]
\drawline\scalar[\E\REG](12000,21000)[3]
\put(12000,21000){\circle*{400}}
\put(12000,18000){\circle*{400}}
\drawline\fermion[\NE\REG](12000,18000)[8400]
\put(18000,24500){k}

\put(20000,21000){+}

\drawline\fermion[\N\REG](23000,16500)[10500]
\drawline\fermion[\N\REG](27500,21000)[6000]
\drawline\scalar[\E\REG](23000,21000)[4]
\put(23000,21000){\circle*{400}}
\put(27500,21000){\circle*{400}}
\put(25500,19500){q}
\put(28000,27000){k}

\put(33000,21000){=}

\drawline\fermion[\N\REG](36500,16500)[10500]
\drawline\fermion[\N\REG](41000,21000)[6000]
\drawline\scalar[\E\REG](36500,21000)[4]
\put(36500,21000){\circle*{400}}
\put(41000,21000){\circle*{1000}}
\put(41500,27000){k}

\drawline\scalar[\W\REG](3000,4500)[2]
\drawline\scalar[\E\REG](6000,4500)[2]
\drawline\fermion[\E\REG](3000,4700)[3000]
\drawline\fermion[\E\REG](3000,4300)[3000]
\drawline\fermion[\N\REG](6000,4500)[6000]
\drawline\fermion[\NE\REG](3000,4500)[8400]
\put(3000,4500){\circle*{400}}
\put(6000,4500){\circle*{400}}
\put(6500,10500){k$_1$}
\put(9500,10500){k}

\put(12000,7500){+}

\drawline\scalar[\E\REG](18000,4500)[2]
\drawline\scalar[\W\REG](18000,4500)[2]
\drawline\fermion[\N\REG](18000,4500)[6000]
\drawline\fermion[\NE\REG](18000,7500)[4200]
\put(18000,4500){\circle*{400}}
\put(18000,7500){\circle*{400}}
\put(18500,10500){k$_1$}
\put(21500,10500){k}

\put(23000,7500){+}

\drawline\fermion[\N\REG](27500,4500)[6000]
\drawline\fermion[\N\REG](32000,4500)[6000]
\drawline\scalar[\W\REG](27500,4500)[1]
\drawline\scalar[\E\REG](27500,4500)[3]
\drawline\scalar[\E\REG](32000,4500)[1]
\put(27500,4500){\circle*{400}}
\put(32000,4500){\circle*{400}}
\put(28000,10500){k$_1$}
\put(32500,10500){k}

\put(35500,7500){=}

\drawline\fermion[\N\REG](40000,4500)[6000]
\drawline\fermion[\N\REG](44500,4500)[6000]
\drawline\scalar[\W\REG](40000,4500)[1]
\drawline\scalar[\E\REG](40000,4500)[3]
\drawline\scalar[\E\REG](44500,4500)[1]
\put(40000,4500){\circle*{400}}
\put(44500,4500){\circle*{1000}}
\put(40500,10500){k$_1$}
\put(45000,10500){k}

\put(24000,500){Fig. 1}
\end{picture}

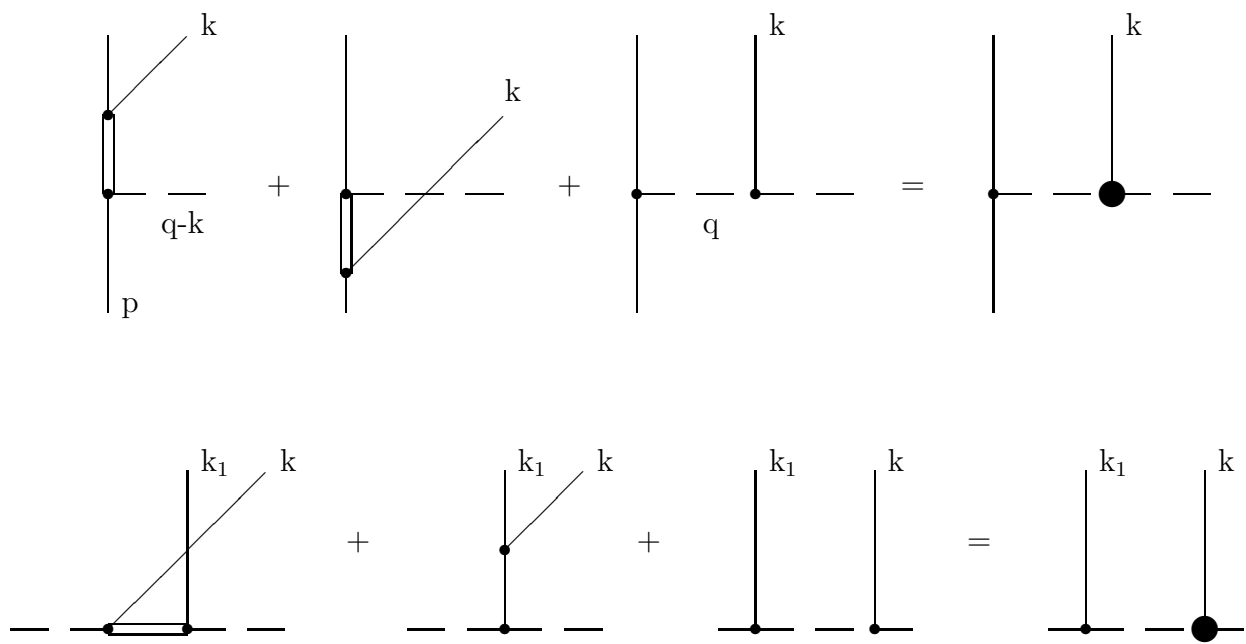
\end{document}